# Utilization of Naturally Occurring Dyes as Sensitizers in Dye Sensitized Solar Cells.

Nipun Sawhney [1], Soumitra Satapathi [1,*]

[1] Department of Physics, Indian Institute of Technology Roorkee, Roorkee, Uttarakhand, 247667, India

## Abstract

Dye sensitized Solar cells (DSSCs) were fabricated with four naturally occurring anthocyanin dyes extracted from naturally found fruits/ juices (viz. Indian Jamun, Plum, Black Currant and Berries) as sensitizers. Extraction of anthocyanin was done using acidified ethanol. The highest power conversion efficiencies ($\eta$) of 0.55% and 0.53% were achieved for the DSSCs fabricated using anthocyanin extracts of blackcurrant and mixed berry juice. Widespread availability of these fruits/juices, high concentration of anthocyanins in them and ease of extraction of anthocyanin dyes from these commonly available fruits render them novel and inexpensive candidates for solar cell fabrication.

## Keywords

# 1. Introduction

Dye sensitized solar cells (DSSCs) or Gratzel cells as discovered by Gratzel et.al.in 1991 have attracted considerable research interest due to their low cost, ease of fabrication and environmental friendliness [1,2]. A DSSC is composed of a porous layer of titanium dioxide (TiO$_2$) coated photoanode, a monolayer of dye molecules that absorbs sunlight, an electrolyte for dye regeneration and a cathode. They form a sandwich like structure with the dye molecule or photosensitizer playing a pivotal role through its ability to absorb visible light photons. Therefore, a significant amount of research on DSSC has been focused on designing and optimizing the photosensitizer in order to absorb a wide spectrum of wavelengths and increase the efficiency of the solar energy conversion [3,4].

Initially ruthenium based complexes had received particular interest as photo sensitizers due to their favorable photo electrochemical properties and high stability in the oxidized state [5]. Ruthenium(II) based dyes in conjunction with iodide-based electrolytes have successfully been employed to achieve maximum power conversion efficiency (PCE) of 11.9% [6]. However the scarcity of noble metals and high cost of Ru dyes (>$1,000/g) limits their large scale commercialization [7].

As a viable alternative, organic dyes with D-π-A or push-pull architecture were designed for improving short circuit current density [8]. Molecular engineering of porphyrin dyes by Simon Mathew, et. al. has resulted in a record 13% efficiency, while a metal-free all organic DSSC with the efficiency of 12.8 % has recently been reported [9,10]. However, synthesis and design of these dyes are complex and they pose high environmental and health risks due to their non-biodegradability and carcinogenic nature [11,12]. Nonetheless, replacement of organic dyes with eco-friendly, biodegradable and cost effective natural dyes opens up a new direction for the commercialization of this technology [13]. Natural dye extracts from vegetables such as Red Turnip and Pomegranate have been employed to obtain power conversion efficiencies of 1.7% and 1.5% [13]. Though the overall efficiency of natural dye based DSSCs still remains low, recent work on structural modification has allowed for performance as good as, if not better than their synthesized counterparts [14]. Molecular engineering of Coumarin dyes by Wang et al. [14] resulted in high efficiencies of 7.6% and 6%.

One of the most abundant and widespread groups of natural pigments are anthocyanins. They are natural dyes that are responsible for coloration of a large number of fruits, leaves and plants [15]. Due to carbonyl and hydroxyl groups present on the anthocyanin molecule, they can bind easily with the surface of TiO$_2$ nanoparticles [16].

In this paper we report the extraction of natural dyes (specifically anthocyanins), from widely available fruit sources and their subsequent use in DSSCs.

# 2. Experimental

### 2.1. Materials

FTO coated Glass, I$^-$/I$_3^-$ Electrolyte and Pt Paste were purchased from Dye Sol, Australia. Titanium (IV) isopropoxide, ethanol, acetone, 35% hydrochloric acid were purchased from HiMedia(Mumbai). jamun (Syzygium Cumini) was plucked from trees growing in the campus of Indian Institute of Technology, Roorkee, while plum (Prunus Salicina) was purchased from local market. Berry juice was obtained from Dabur (India) and black currant (Ribes Nigrum) pulp was obtained from Mala's (Mumbai).

## 2.2. Methods

### 2.2.1. Preparation of Anthocyanin Fruit Extracts.

Fresh jamuns were plucked from the trees of IIT Roorkee. The skin and flesh were separated from the seeds using a spatula. White fleshed jamuns were separated as their color indicated lower concentrations of anthocyanins in them [17]. The deep purple colored jamuns were used for this study as they have higher anthocyanins concentration [17]. Fresh plums were purchased from the local market keeping in mind the requirement of deeper coloration so as to maximize anthocyanins concentration in the extract. Skin and flesh were then separated from the seeds using a spatula. Seeds of these fruits were then disposed due to low/negligible concentration of anthocyanins [18]. Black currant pulp and mixed berry juice were purchased from the local market and were used as received. Dyes were extracted from jamun, plum, black currant and mixed berry juice by a modified procedure similar to the procedure reported earlier [18]. One cup of skin and pulp was grinded by hand for 2 hours using mortar and pestle and then sonicated in 75% ethanol acidified with 10 mM HCl for one hour. The mixture was then centrifuged at 1500 rpm for 10 minutes and decanted. Large particles were removed from the solution using a stainless steel mesh and the solution was filtered using a filter paper.

### 2.2.2 Fabrication of DSSCs

FTO slides were successively cleaned in distilled water, ethanol and acetone for cycles of 30 minutes each respectively. Synthesis of nano-crystalline $TiO_2$ was done through hydrolysis of titanium (IV) isopropoxide [19]. Titanium (IV) isopropoxide was mixed with ethanol, and the mixture was stirred at 800 rpm for 30 minutes on a magnetic stirrer. This was followed by ultrasonicated for 30 minutes. 24 ml distilled water was then added at pace of nearly .5ml/minute for 48 minutes using a micropipette. The mixture was then heated to 353 K under reduced pressure to remove the solvent. It was then further dried in an oven at 393 K for a few hours to obtain dry $TiO_2$ powder. $TiO_2$ powder was then calcined in a muffle furnace at 673K to produce anatase $TiO_2$ nanoparticles [19]. $TiO_2$ slurry was then obtained by slowly adding pH3 acetic acid solution. The slurry was ultrasonicated for 30 minutes, and was placed on a magnetic stirrer at 1200 rpm for 30 minutes. Ultrasonication and stirring was repeated 4 times to get consistent viscous slurry of $TiO_2$.

The prepared $TiO_2$ slurry was then doctor bladed on FTO glass slide using doctor tape (30 um) on one side and sintered at an optimized temperature of 550K [20]. The photo anodes were dipped in the respective dye solutions for 12 hrs for complete dye loading. Pt counter electrodes were prepared using Pt paste as per the procedure given in [21]. The electrodes and counter electrodes were used to make sandwiched devices with active cell area of $0.15 cm^2$.

### 2.2.3. Measurements

Scanning Electron Microscopy images of electrodes were taken after sputtering the surface with gold to make it conducting using Scanning Electron Microscope (JEOL JEM2010) operated at 15kV and magnification of 2000X.

Cells corresponding to Black Currant and Mixed Berry Dye extracts were tested under 350 W/m$^2$ illuminations using artificial sunlight while cells corresponding to Jamun and Plum Dye extracts were tested under 960 W/m$^2$ using Zahner's CIMPS-QE/IPCE system and Newport CIMPS-QE/IPCE system respectively, coupled with Oriel Sol3A Class AAA Solar Simulator with 1000 Watt Xenon lamp (model number 94063A).

UV-Visible absorption spectra of dye solutions were taken after dilution in ethanol and subsequent ultrasonication using Hitachi UV-Vis Spectrophotometer with the range of 200 nm to 1100 nm.

## 3. Results and Discussions

Fig. 1(a) and Fig. 1(b) show the Scanning Electron Microscopy (SEM) images of a bare, sintered $TiO_2$ layer and the anthocyanin dye loaded $TiO_2$ layer respectively. Highly porous electrode layers with large surface areas are readily available to dyes for rapid adsorption, thus enhancing the short circuit current density. As a result mesoporous electrode materials show enhanced electrochemical performance compared with their bulk counterparts [22]. The presence of holes and extensive bridging as depicted in Fig. 1(a) is an indicator of highly conducting mesoporous and crystalline $TiO_2$ layers whereas Fig. 1(b) reveals extensive adsorption of the anthocyanin dye on the $TiO_2$ nanoparticles.

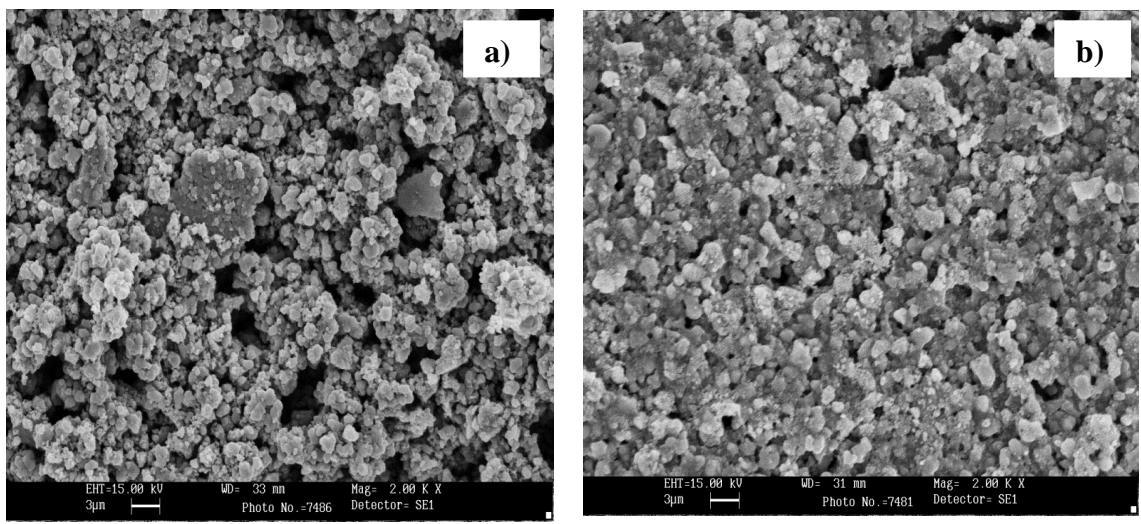

Fig. 1.

SEM images of (a) bare $TiO_2$ layer and (b) anthocyanin dye loaded $TiO_2$ layer.

The structure of a typical anthocyanin molecule and its binding with $Ti^{4+}$ is depicted below in Fig. 2. [16]

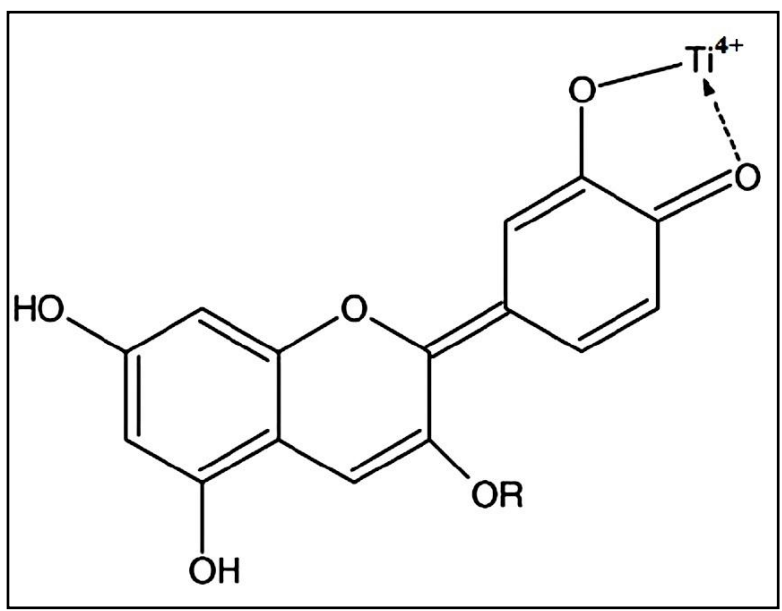

Fig. 2
Structure and binding of a typical anthocyanin molecule with $Ti^{4+}$.

Fig. 3 shows the UV-Visible absorption spectra of different sensitizers used in this study. As shown in the figure, the UV-visible absorption maximum occurs at 538 nm for Jamun. The Plum extract shows two bands in its absorption spectrum viz. at 536 nm and 538 nm which could be attributed to the presence of delphinidin derivatives specifically delphinidin 3-glucoside, petunidin 3-glucoside and malvidin 3-glucoside [23, 24]. The UV-vis absorption maximum at 526 nm for black currant could be attributed to the presence of 3-galactosides of malvidin and delphidin [23, 24]. The mixed berry extracts has an absorption maximum at 514.5 nm which could be attributed to the cyanidin derivatives (cyanidin 3-glucoside, -galactoside, and –arabinoside) [23, 24]. Since the anthocyanin isomers were not isolated before taking their absorption spectra; therefore instead of sharp absorption peaks, broader peaks were observed which are indicative of a mixture of isomers. However, the characteristics features of the absorption spectra between 510 and 540 nm are indicative of the presence of high concentration of anthocyanin isomers in the above extracts [23-27].

Fig. 4 shows the incident photon to current efficiency (IPCE) spectrum of DSSC device sensitized with Jamun dye.

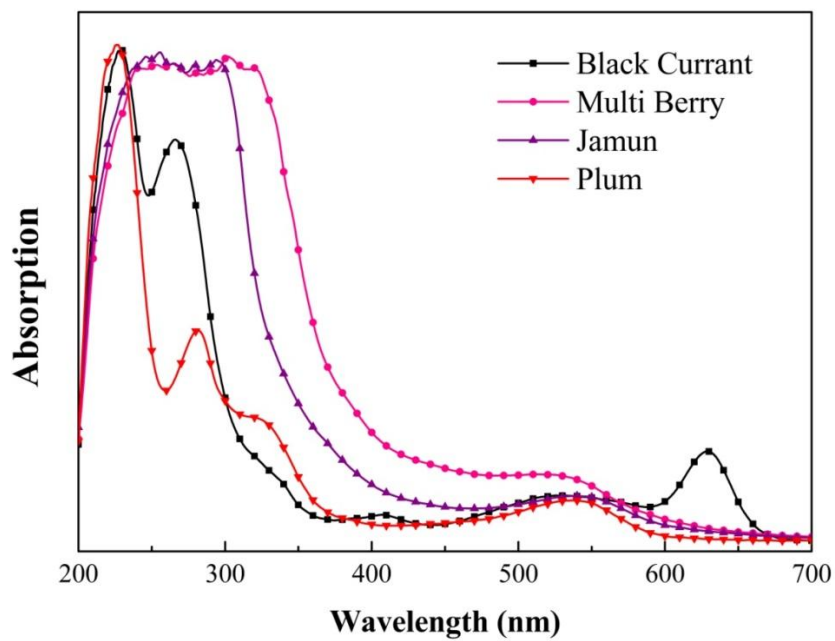

Fig. 3.
UV-Visible absorption spectra of the fruit extracts in ethanol

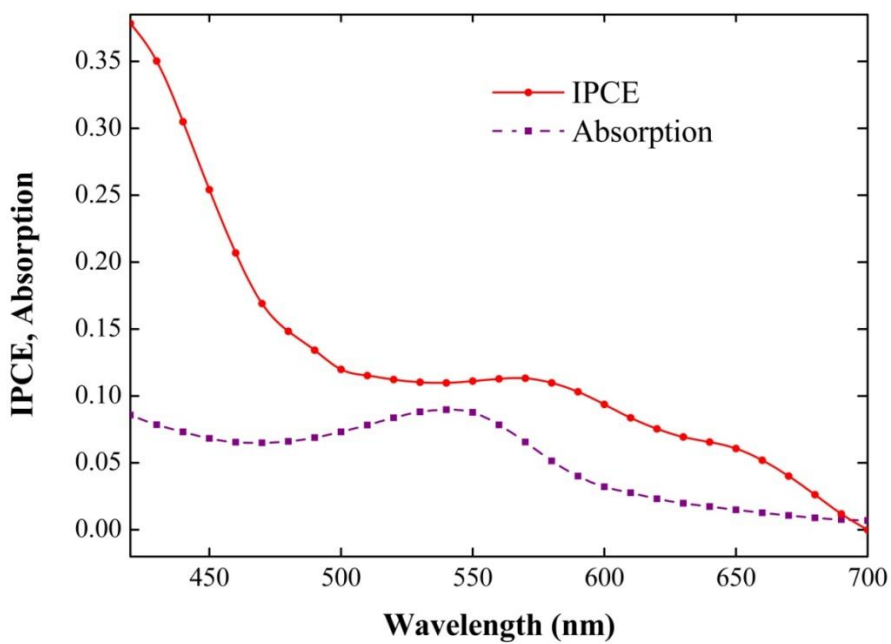

Fig. 4.
Comparison of the absorption spectra of Jamun dye extract and IPCE of the Jamun adsorbed on TiO$_2$ as in the Dye Sensitized Solar Cell.

Fig. 5 depicts the I-V characteristics of the DSSCs made from extracts of black currant crush and berry juice. Highest power conversion efficiency (η) of 0.55% and 0.53% were obtained for these cells with $V_{oc}$ of 0.536 V and 0.522 V, $J_{sc}$ of 6.417 A/m$^2$ and 6.081 A/m$^2$ and Fill Factor (FF) of 0.56 and 0.58 respectively. The corresponding values are given in Table [1]. Fig. 6 depicts the I-V characteristics of the DSSC devices made from the extracts of freshly plucked jamun (Syzygium Cumini) and plum (Prunus Salicina). They resulted in efficiencies of 0.23% and 0.26% respectively with $V_{oc}$, $J_{sc}$ and FF values as shown in Table [2]. The short circuit current density ($J_{sc}$) of DSSCs depends on the intensity of the incident light. On varying the intensity of illumination from 350 Wm$^{-2}$ to 960 Wm$^{-2}$, an increase of $J_{sc}$ was observed from 6.08 to 23.62 Am$^{-2}$ for mixed berry and plum respectively. Although, the $J_{sc}$ increases with increasing incident light intensity, the overall power conversion efficiency (η) of plum sensitized DSSC is lower than that for mixed berry due to lower stability of anthocyanin based DSSC at high illumination intensity.

Dyes extracted from black currant crush have resulted in the DSSCs with highest efficiency. This is expected because of the presence of deeper and darker coloration of the fruit due to the presence of anthocyanins which have an increased absorption of light in visible spectra [28]. The result leads us to believe that darker dyes are preferable candidates due to their increased absorption of visible light as compared to lighter ones leading to a increased photo current densities. Moreover anthocyanin concentration in black currant (>5mg/gram) and various berries (>3.5 mg/gram) is significantly higher than the concentration in plum (<1mg/gram) and jamun (<2.5 mg/gram) [17, 29]. This allows for increased purity of the extract due to higher relative concentrations of anthocyanins compared to impurities (such as other phenolics), which is responsible for increased dye loading and thereby increased efficiencies.

**Table 1**
I-V Characteristics of Black Currant and Mixed Berry based solar cells at 350W/m$^2$ illumination.

| S. No. | Dye Source | Voc (V) | Jsc (A/m$^2$) | Fill Factor | Efficiency (%) |
|---|---|---|---|---|---|
| 1 | Black Currant | 0.536 | 6.417 | 0.559 | 0.55 |
| 2 | Mixed Berry | 0.522 | 6.081 | 0.583 | 0.53 |

**Table 2**
I-V Characteristics of Jamun and Plum based solar cells at 960W/m$^2$ illumination

| S. No. | Dye Source | Voc (V) | Jsc (A/m$^2$) | Fill Factor | Efficiency (%) |
|---|---|---|---|---|---|
| 1 | Jamun | 0.403 | 14.589 | .383 | .235 |
| 2 | Plum | 0.352 | 23.615 | .301 | .261 |

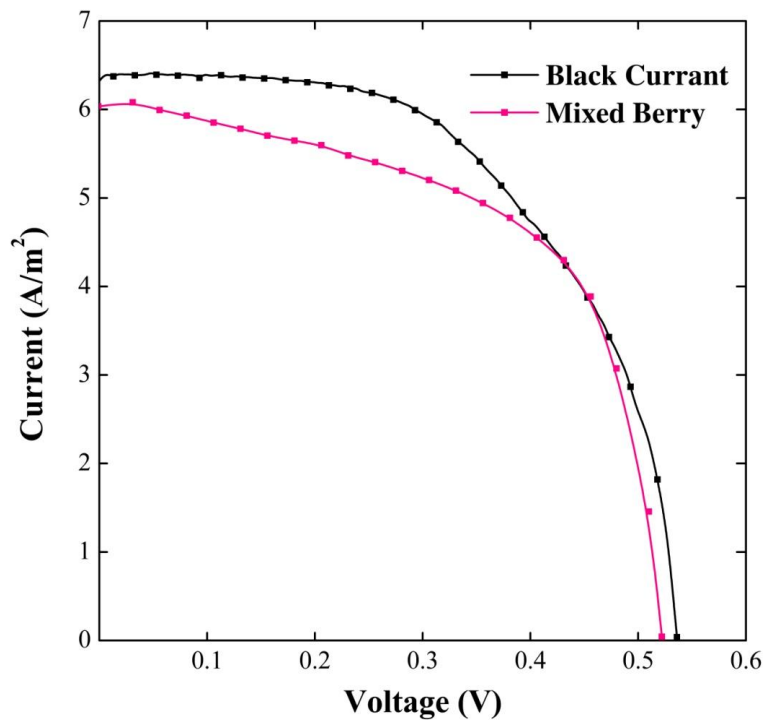

Fig. 5

I-V Characteristics of DSSCs made using Black Currant and Mixed Berry Extracts, taken under 350 W/m$^2$ illumination of artificial sunlight.

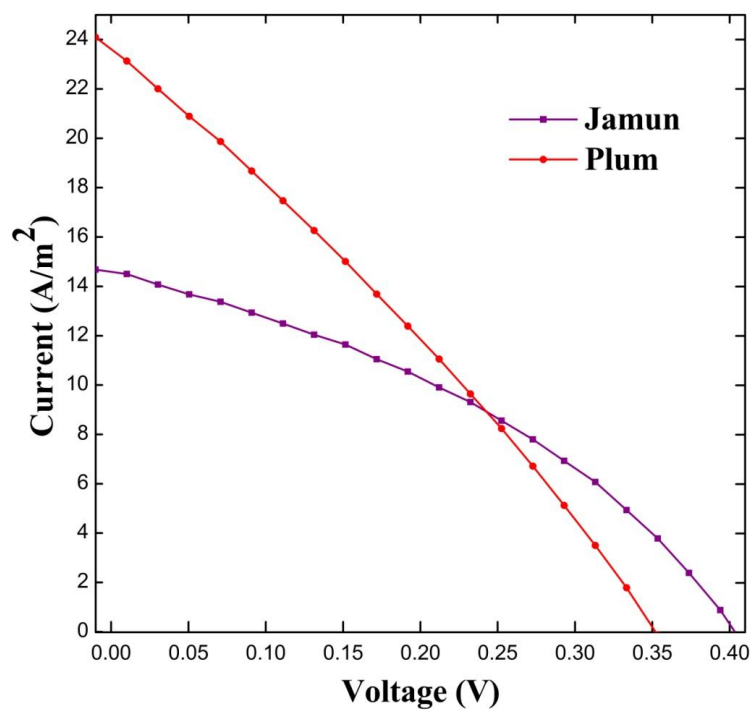

Fig.6

I-V Characteristics of DSSCs made using Jamun and Plum extracts taken under 960 W/m$^2$ illumination of artificial sunlight.

## 4. Conclusions

In summary, we have investigated the utilization of naturally occurring dyes, specifically anthocyanins extracted from various commonly found fruits, using the economical and efficient procedure for DSSC fabrication. Dyes extracted from black currant crush and mixed berries have resulted in the DSSCs with the highest efficiencies. It has been observed that darker colored anthocyanins lead to higher efficiency. This is because darker color corresponds to increased light absorption leading to enhanced photo current densities. Moreover black currant crush and mixed berry had higher concentrations of anthocyanins leading to increased purity of extracts, which resulted in better dye loading. However, the lower current densities in these dyes as compared to commercial Ru based dyes could be attributed to the additional impurities resulting from imprecise extraction processes. The isolation and purification of the various isomers will address this issue and could potentially improve the power conversion efficiency. The simplicity and cost effectiveness of the overall fabrication process, widespread availability of these fruits/juices, and ease of extraction of anthocyanin dyes from these commonly available fruits render them novel and inexpensive candidates for solar cells application.

## Acknowledgments

SS acknowledges Faculty Initiative Grant (Grant No: ) from Indian Institute of Technology Roorkee.
NS acknowledges Department of Science and Technology Inspire Scholarship.

# References

[1] B. O'regan, M. Grätzel. A low-cost, high-efficiency solar cell based on dye-sensitized colloidal TiO2 films. Nature 353.6346 (1991) 737-740.

[2] M. Grätzel. Photoelectrochemical cells. Nature 414.6861 (2001) 338-344.

[3] B. E Hardin., H.J. Snaith, M.D. McGehee. The renaissance of dye-sensitized solar cells. Nature Photonics 6.3 (2012) 162-169.

[4] S. Satapathi, et al. Performance enhancement of dye sensitized solar cells by incorporating graphene nanosheets of various sizes, Applied Surface Science 314 (2014) 638-641

[5] O. Kohle, et al. The photovoltaic stability of, bis (isothiocyanato) rlutheniurn (II)-bis-2, 2′ bipyridine-4, 4′-dicarboxylic acid and related sensitizers. Advanced Materials 9.11 (1997) 904-906

[6] R. Komiya, et al. Technical Digest, 21st International Photovoltaic Science and Engineering Conference 2 C-5O-08 (2011)

[7] Y. Qin, Q. Peng. Ruthenium sensitizers and their applications in dye-sensitized solar cells. International Journal of Photoenergy 2012 (2012).

[8] R.K. Kanaparthi, J. Kandhadi, L. Giribabu. Metal-free organic dyes for dye-sensitized solar cells: recent advances. Tetrahedron 68.40 (2012) 8383-839.

[9] S. Mathew, et al. Dye-sensitized solar cells with 13% efficiency achieved through the molecular engineering of porphyrin sensitizers. Nature chemistry 6.3 (2014) 242-247.

[10] K. Kakiage, et al. Fabrication of a high-performance dye-sensitized solar cell with 12.8% conversion efficiency using organic silyl-anchor dyes. Chemical Communications 51.29 (2015) 6315-6317.

[11] A.M.C. Batlle. Porphyrins, porphyrias, cancer and photodynamic therapy—a model for carcinogenesis. Journal of Photochemistry and Photobiology B: Biology 20.1 (1993) 5-22.

[12] K.T. Chung. The significance of azo-reduction in the mutagenesis and carcinogenesis of azo dyes. Mutation Research/Reviews in Genetic Toxicology 114.3 (1983) 269-281.

[13] M.R. Narayan. Review: dye sensitized solar cells based on natural photosensitizers. Renewable and Sustainable Energy Reviews 16.1 (2012) 208-215.

[14] Z.S. Wang, et al. Molecular design of coumarin dyes for stable and efficient organic dye-sensitized solar cells. The Journal of Physical Chemistry C 112.43 (2008) 17011-17017.

[15] Ø.M. Andersen and M. Jordheim. Anthocyanins. eLS (2006).

[16] S. Hao, et al. Natural dyes as photosensitizers for dye-sensitized solar cell. Solar energy 80.2 (2006) 209-214.

[17] J.M. Veigas, et al. Chemical nature, stability and bioefficacies of anthocyanins from fruit peel of Syzygium cumini Skeels. Food Chemistry 105.2 (2007) 619-627.

[18] F. Aqil, et al. Antioxidant and antiproliferative activities of anthocyanin/ellagitannin-enriched extracts from Syzygium cumini L.(Jamun, the Indian Blackberry). Nutrition and cancer 64.3 (2012) 428-438.

[19] R.J. Tayade, R.G. Kulkarni, R.V. Jasra. Photocatalytic degradation of aqueous nitrobenzene by nanocrystalline TiO2. Industrial & engineering chemistry research 45.3 (2006) 922-927.

[20] S. Schattauer, et al. Influence of sintering on the structural and electronic properties of TiO2 nanoporous layers prepared via a non-sol–gel approach. Colloid and Polymer Science 290.18 (2012): 1843-1854.

[21] S.H. Kim, C. W. Park. Novel Application of Platinum Ink for Counter Electrode Preparation in Dye Sensitized Solar Cells. Bull. Korean Chem. Soc 34.3 (2013) 831.

[22] Y.G. Guo, J.S. Hu, and L.J. Wan. Nanostructured materials for electrochemical energy conversion and storage devices. Adv. Mater 20.15 (2008) 2878-2887.

[23] V. Hong, R.E. Wrolstad. Use of HPLC separation/photodiode array detection for characterization of anthocyanins. Journal of Agricultural and Food Chemistry 38.3 (1990) 708-715.

[24] J.B. Harborne. "Spectral methods of characterizing anthocyanins." Biochemical Journal 70.1 (1958): 22.

[25] R. Slimestad, and H. Solheim. Anthocyanins from black currants (Ribes nigrum L.). Journal of Agricultural and Food Chemistry 50.11 (2002) 3228-3231.

[26] A.F. Faria, M.C. Marques, and A.Z. Mercadante. Identification of bioactive compounds from jambolão (Syzygium cumini) and antioxidant capacity evaluation in different pH conditions. Food chemistry 126.4 (2011) 1571-1578.

[27] M.M. Giusti, R. E. Wrolstad. Characterization and measurement of anthocyanins by UV-visible spectroscopy. Current protocols in food analytical chemistry (2001).

[28] G. Wyszecki, W. S. Stiles. Color science. Vol. 8. New York: Wiley, 1982.

[29] X. Wu, et al. Concentrations of anthocyanins in common foods in the United States and estimation of normal consumption. Journal of Agricultural and Food Chemistry 54.11 (2006) 4069-4075.